\documentclass[12pt]{article}
\usepackage{amssymb}
\usepackage{amsfonts}
\usepackage{amsmath}
\usepackage{amsmath,amsthm}
\usepackage{subfigure}
\usepackage{graphicx,epsfig, color }

\oddsidemargin 10mm \numberwithin{equation}{section}
\begin{document}
\begin{center}
{{\bf {Galaxy rotation curves and preferred reference frame
effects}}\vskip 1 cm { Hossein Ghaffarnejad
       \footnote{E-mail address: hghafarnejad@semnan.ac.ir} and Razieh  Dehghani
       \footnote{E-mail address: Razieh-Dehghani@semnan.ac.ir}}}\\
   \vskip 0.1 cm
   {\textit{Faculty of Physics, Semnan
            University, Semnan, 35131-19111, IRAN}} \\

    \end{center}
\begin{abstract}
     As an alternative to dark matter models we use generalized Jordan-Brans-Dicke scalar-vector-tensor (JBD-SVT) gravity model to study
     the behavior  of the rotational velocities of test particles moving around galaxies. To do so we consider an interaction
     potential $U(\phi,N_{\mu})$ between the Brans-Dicke scalar field $\phi$ and time like dynamical four-vector field $N_\mu$ which plays as
     four velocity of a preferred reference
     frame. We show that at in weak field limits metric solution
     of the galaxy under consideration reaches to a modified
     Schwarzschild-de Sitter space in which mass of the vector
     field plays as an effective cosmological constant.
     In fact the present work proposes modification on the formulation of Newton's gravitational acceleration.
       This is used to explain circular
      velocity of galaxies without postulating dark matter. We also check our theoretical results with
      empirical baryonic Tully Fisher relation which states  a linear relations between the rotational speed of galaxies and their
      mass.
      Mathematical calculation predict a good correspondence between our theoretical results and experimental observations
      for a set of 12 spiral galaxies.
\end{abstract}
\section{Introduction}

Observations of the dynamics of galaxies as well as the dynamics
of the whole Universe indicates that a main part of the Universe's
mass must be missing. According to the works done by authors in
Ref. \cite{GB}, this missing mass is possibly made by (the
unknown)  dark matter. So galactic scale dynamics is one of the
important systems which is subjected to the dark matter studies.
The observations of galaxies show that there is a discrepancy
between the observed dynamics and the mass inferred from luminous
matter \cite{Rub,Ford}. Since no dark matter has been detected so
far, replacing dark matter by a modified gravity theory, is an
alternative approach to the problem of missing mass \cite{JM}.
There are different approaches to solve this problem, such as
Modified Newtonian Dynamics (MOND) \cite{RH} and it's relativistic
extension \cite{Bek}, Modified Gravity (MOG) or its generally
covariant version that called scalar-vector-tensor gravity
theory(STVG) \cite{JM}, conformal gravity \cite{MK}, nonsymmetric
gravity theory (NGT) \cite{JCAP} and nonlocal gravity
\cite{mash2007,mash2009}.
  In
STVG theory the dynamics of a test particle is given by a modified
equation of motion. Since the metric field is coupled to scalar
fields and a massive vector field, the solution of the field
equations for a point mass is different from the point mass
Schwarzschild solution of general relativity \cite{MT}. The
predictions of STVG theory for the rotation curves of galaxies
have been compared to observational data \cite{BM}, by using a
static spherically symmetric point mass metric solution for
galaxies under consideration. The same approach has also been
applied to the dynamics of globular clusters and clusters of
galaxies \cite{BM2,MT2}. \ \
 According to the works done by
authors in refs. \cite{Moffat1, Moffat2, CLJM, Barbero, Jacob,
Barbero2} a new scalar-vector-tensor gravity theory is introduced
where a time-like dynamical vector field can change a Lorentzian
signature of the background to an Euclidian form. The dynamical
vector field is usually considered as four velocity of preferred
reference frames.
 Another approach
was used to extend the Jordan-Brans-Dicke scalar-tensor gravity
theory \cite{BD} and called as Jordan-Brans-Dicke
scalar-vector-tensor (JBD-SVT) model by one of the authors in
refs. \cite{GH1, GH2}. Several applications of this model are
studied in refs.\cite{GH3,GH4,GH5,GH6}.
 The main motivations to present this JBD-SVT gravity model are as follows:
The exact Lorentz invariance is impossible to test uniformly, as
the boost parameter of this group is unrestrained. It also leads
to the problem of divergences in quantum field theory associated
with states of arbitrarily high energy and momentum \cite{GH1}.
This problem can be solved by a short distance high energy cutoff
length which violates the Lorentz invariance \cite{jac}. Lorentz
invariance violation causes to change the metric signature of
space-time. For these reasons, we scrutinize the possibility that
there is a preferred rest frame at each space-time point. If this
frame was to be a fixed structure, it would violate general
covariance, then the matter energy-momentum tensor is not
divergence-less and the Einstein field equations is inconsistent.
In order to preserve general covariance, the preferred frame
should be considered dynamical. In this case dynamical frame could
be defined by a vector field \cite{CM1,CM2,RW,MA} or by the
gradient of a scalar field \cite{MA2,BA}. One of the remarkable
results of a suitable dynamical preferred frame, is to have two
metrics with inequivalent causal structure named as Lorentzian and
Euclidean metrics, respectively. The quantum cosmology viewpoint
of the very early universe models describe its origin as a quantum
tunneling from Euclidian to the Lorentzian space-time \cite{GH1}.
In both of them the space-time coordinates are still real
coordinates. Whereas an ordinary way to obtain an Euclidean metric
solution of the Einstein field equations, is to introduce a
complex time variable $\tau = it$, so the Lorentzian signature of
the metric is changed to its Euclidean version \cite{GH1}.\\
 In accordance with the works
has been done previously \cite{JM, MK, JCAP, mash2007, mash2009},
in the weak field approximation the potential for a matter
distribution of an extended object behaves like the Newtonian
potential with an enhanced gravitational constant and an
additional Yukawa potential. Applying a general potential
$U(\phi,N_\mu)$ and weak field approximation of the background
metric obtained from dynamical field equations of the JBD-SVT
gravity model \cite{GH1} we study rotation curves of galaxies
without postulating exotic dark matter which usually is described
by dynamical scalar fields.
 This paper is organized as follows:\\
In section 2 we defined the JBD-SVT gravity \cite{GH1} and
calculate weak field limit of the Lagrangian density of the model
for spherically symmetric state metric equation. Then we obtained
linear order solutions of the Euler-Lagrange equations of the
fields. In section 3 the acceleration equation and rotational
velocity of a test particle is calculated in the weak field limit.
Section 4 is devoted to the observational tests of the model. It
is shown that the model is consistent with observational rotation
curves of a sample of 12 spiral galaxies and the empirical
Tully-Fisher relation.
 Last section denotes to
concluding remarks and outlook of the work.
 \section{The model}
Let us start with the following JBD-SVT gravity model \cite{GH1}
\begin{equation}\label{acf}
 I_{T}=I_{BD}+I_{N}
\end{equation} in which
 \begin{equation}\label{ac11}
  I_{BD}=\frac{1
}{16\pi}\int d^4x\sqrt{-g}\left\{\phi R-\frac{\omega}{
    \phi}
g^{\mu\nu}\nabla_{\mu} \phi\nabla_{\nu}\phi \right\}
\end{equation}
   is JBD scalar-tensor
gravity \cite{BD}. $g$ is absolute value of determinant of the
metric tensor $g_{\mu \nu}$ where we use the metric signature
convention as (-,+,+,+).  $\phi$ is the Brans-Dicke scalar field
and $\omega$ is the Brans-Dicke adjustable coupling constant. The
second term of the action (2.1) is
 \begin{equation}\label{ac2}
 I_N=\frac{1}{16\pi}\int
d^4x\sqrt{-g}\{\zeta(x^{\nu})(g^{\mu\nu}N_{\mu}N_{\nu}+1)+2\phi
F_{\mu\nu}F^{\mu\nu}+U(\phi,N_{\mu})$$$$-\phi N_\mu
N^{\nu}(2F^{\mu\lambda}\Omega_{\nu\lambda}+
F^{\mu\lambda}F_{\nu\lambda}+\Omega^{\mu\lambda}\Omega_{\nu\lambda}
-
2R^{\mu}_{\nu}+\frac{2\omega}{\phi^2}\nabla^{\mu}\phi\nabla_{\nu}\phi)\},
\end{equation}
for which
\begin{equation}\label{fo}
F_{\mu\nu}=2(\nabla_{\mu}N_{\nu}-\nabla_{\nu}N_{\mu}),~~~~~~~\Omega_{\mu\nu}=2(\nabla_{\mu}N_{\nu}+\nabla_{\nu}N_{\mu})
\end{equation}
describes action of a unit time-like dynamical four vector field
$N_\mu.$ In other words it is the action of a preferred reference
frame with four velocity $N_{\mu}(x^{\nu})$. Up to the terms of
$\zeta(x^{\nu})$ and the scalar potential $U(\phi,N_{\mu})$, the
action (\ref{ac2}) is obtained by transforming the JBD action
(\ref{ac11}) under the metric transformation $g_{\mu\nu}\to
g_{\mu\nu}+2N_{\mu}N_{\nu}.$ One can follow ref. \cite{GH1} for
more details. Action (\ref{ac2}) shows that the vector field
$N_{\mu}$ is coupled non-minimally to the JBD scalar field $\phi$
and the metric field $g_{\mu\nu}.$ Action (\ref{acf}) is written
in units $c=G=\hbar=1$ . The undetermined Lagrange multiplier
$\zeta(x^{\nu})$ controls $N_{\mu}$ to be an unit time-like vector
field. $\phi$ describes inverse of variable Newton's gravitational
coupling parameter and its dimension is $(lenght)^{-2}$ in units
$c=G=\hbar=1$. Present limits of dimensionless BD parameter
$\omega$ based on time-delay experiments \cite{CW1,CW2,EG,RD}
requires $\omega\geq4\times10^{4}.$ General relativistic approach
of the BD gravity action (\ref{ac11}) is obtained by setting
$\omega\to\infty$ \cite{Cho} for which one can infer
\begin{equation}\label{om}
\lim_{\omega\to\infty}\phi\approx\frac{1}{G}
\end{equation}
 where $G$ is the Newtonian coupling constant.\\
 Let us consider a general
spherically symmetric static metric equation which can be
expressed in the following isotropic form \cite{BD}
\begin{equation}\label{meta}
ds^2=-e^{2\epsilon\alpha(r)}dt^2+e^{2\epsilon\beta(r)}[dr^2+r^2
d\theta^2+r^2\sin^2\theta d\varphi^2]
\end{equation}
 where $\epsilon$ is a constant.
In order to study the behavior of JBD-SVT in astrophysical scales
we should apply weak field approximation for the
  dynamics of the fields by perturbing the fields around Minkowski spacetime. In this case
 $\epsilon$ should have small values and will be order parameter of the perturbation. Furthermore, $\epsilon$
  should be defined versus the parameter of the model which comes from non-linear counterpart of the action.
 In the BD action (\ref{ac11}), $\omega$ determines
  nonlinear counterpart of the action. Hence
  it is suitable to choose $\epsilon=\epsilon(\omega).$
   One can write perturbation series form of the fields which up to the second order terms are
    \begin{equation}\label{sera}
     e^{2\epsilon\alpha(r)}\approx1+2\epsilon
\alpha(r)+O(\epsilon^2)
\end{equation}
\begin{equation}\label{serb}
e^{2\epsilon\beta(r)}\approx1+2\epsilon\beta(r)+O(\epsilon^2),
\end{equation}
and
\begin{equation}\label{serp}
\phi(r)\approx\frac{1}{G}\bigg[1+\epsilon
\psi(r)+O(\epsilon^2)\bigg].
\end{equation}
 For static
spherically symmetric metric (\ref{meta}) the vector field $N_\mu$
and the tensor fields $F_{\mu\nu}$ and $\Omega_{\mu\nu}$ should
depend on the radial coordinate $r$ for which one can choose
\cite{DG1}
\begin{equation}\label{vec}
N_{\mu}(r)=(b(r), q(r),0,0)
\end{equation} where the time like condition $g^{\mu\nu}N_\mu N_\nu=-1$ in weak field limits reads
 \begin{equation}\label{timelike}
 -b^2(r)+q^2(r)\approx-1.
\end{equation}
Substituting (\ref{vec}) into (\ref{fo}) we obtain
\begin{equation}\label{fd}
F_{tr}(r)=-2b^{\prime}(r),~~~\Omega_{tr}(r)=2b^{\prime}(r)-4\epsilon
b(r)\alpha^{\prime}(r),
\end{equation}
 and
\begin{equation}\label{omd}
~~~\Omega_{tt}\approx-4\epsilon
q\alpha^{\prime},~~~\Omega_{rr}\approx 4q^{\prime}-4\epsilon
q\beta^{\prime}
\end{equation}
where prime denotes to derivative with respect to $r$ coordinate
and we used linear order terms of non-vanishing Christoffel
symbols $\Gamma_{tt}^r(r)=\Gamma^t_{tr}(r)\approx \epsilon
\alpha^{\prime}(r)$ and $\Gamma_{rr}^r\approx
\epsilon\beta^{\prime}(r)$ for small $\epsilon$. Substituting
(\ref{sera}) and (\ref{serb}) we obtain $tt$ and $rr$ components
of the Ricci tensor for small $\epsilon$ as
\begin{equation}\label{elem3}
R_{tt}\approx-\frac{\epsilon}{r}[r\alpha^{\prime\prime}+2\alpha^{\prime}]+O(\epsilon^2),~~~R_{rr}\approx\frac{\epsilon}{r}[2\beta^{\prime}+
r\alpha^{\prime\prime}+2r\beta^{\prime\prime}]+O(\epsilon^2)
\end{equation}
where the 4D Ricci scalar $R^\mu_\mu$ reads
\begin{equation}\label{elem4}
R\approx\frac{2\epsilon}{r}[2\alpha^{\prime}+4\beta^{\prime}+r\alpha^{\prime\prime}+2r\beta^{\prime\prime}]+O(\epsilon^2)
.\end{equation} Substituting the above perturbative functions into
the BD action (\ref{ac11}) we obtain $I_{BD}=\int dt dr
\mathcal{L}_{BD}$ in which $\mathcal{L}_{BD}$ is the Brans Dicke
Lagrangian density such that
\begin{equation}\label{LBD}\mathcal{L}_{BD}=\frac{\epsilon}{G}\left[r\alpha^{\prime}+2r\beta^{\prime}+\frac{r^2\alpha^{\prime\prime}}{2}
+r^2\beta^{\prime\prime}\right]$$$$
+\frac{\epsilon^2}{G}[(\alpha+3\beta+\psi)[r\alpha^\prime+2r\beta^\prime+\frac{r^2\alpha^{\prime\prime}}{2}+r^2\beta^{\prime\prime}]-\frac{\omega}{4}
r^2\psi^{\prime2}]
\end{equation}
Integrating by part, we remove $\alpha^{\prime\prime}$ and
$\beta^{\prime\prime}$ terms of the Lagrangian density (\ref{LBD})
to obtain its effective counterpart which up to third  order term
$O(\epsilon^3)$ is as follows
\begin{equation}\label{bdeff}
\mathcal{L}_{BD}^{eff}=-\frac{\epsilon^2r^2}{G}\bigg[\frac{\omega\psi^{\prime2}}{4}+\frac{\alpha^{\prime2}}{2}+3\beta^{\prime2}+\frac{5}{2}\alpha^\prime\beta^{\prime}+\frac{\alpha^\prime\psi^\prime}{2}+\beta^\prime\psi^\prime
\bigg].
\end{equation}
 If we use similar calculations for the action functional (\ref{ac2}) then  we will have $I_N=\int
dtdr \mathcal{L}_N$ where $\mathcal{L}_N$ is

the Lagrangian density of the non-minimal interacting vector field
which is defined by
\begin{equation}\label{neff}
\mathcal{L}_N=\mathcal{L}_N^{(0)}+\epsilon\mathcal{L}_N^{(1)}+\epsilon^2\mathcal{L}_N^{(2)}+\cdots
\end{equation}
where
\begin{equation}\label{lzero}
\mathcal{L}_N^{(0)}=\frac{r^2}{G}\bigg(-10b^{\prime2}+\frac{V_0(b)}{4}\bigg),
\end{equation}
\begin{equation}\label{lone}
\mathcal{L}_N^{(1)}=\frac{r^2}{G}\bigg\{\frac{V_1}{4}+\frac{(\alpha+3\beta)V_0}{4}+[14\beta+2\alpha-10\psi+12b^2(\alpha-\beta)]b^{\prime2}
$$$$-\frac{\alpha^{\prime\prime}}{2}+(b^2-1)\beta^{\prime\prime}+\frac{(b^2-1)\beta^{\prime}}{r}-\frac{b^2\alpha^{\prime}}{r}+8b(1-b^2)
b^{\prime}\alpha^{\prime}+4b(3-2b^2)b^{\prime}\beta^{\prime}\bigg\}
\end{equation}
\begin{equation}\label{ltwo}\mathcal{L}_N^{(2)}=\frac{r^2}{G}\bigg\{
\frac{V_2}{4}+\frac{(\alpha+3\beta)V_1}{4}+16b^2(b^2-1)\alpha^\prime\beta^\prime-8(b^2-1)^2\beta^{\prime2}-8b^4\alpha^{\prime2}$$$$
+\frac{(1-b^2)\omega\psi^{\prime2}}{2}+(b^2-1)(\alpha-\beta+\psi)\beta^{\prime\prime}+\bigg[\frac{(\beta-\alpha-\psi)}{2}+2b^2(\alpha-\beta)
\bigg]\alpha^{\prime\prime}$$$$
+\frac{(\alpha-\beta)(b^2-1)\beta^\prime}{r}+\frac{[b^2(\alpha-3\beta)-1]\alpha^\prime}{r}+
[4(\alpha+3\beta)(3\alpha+4\beta)+2\psi(5\alpha+\beta)$$$$
+12b^2(\alpha-\beta)(\alpha+3\beta-\psi)]b^{\prime2}+4[8\beta+3\psi+2b^2(5\beta-3\alpha-\psi)]bb^{\prime}\alpha^{\prime}$$$$+
4[\alpha+29\beta-2\psi+b^2(2\psi-22\beta+2\alpha)]bb^{\prime}\beta^{\prime}
\bigg\}.
\end{equation}
Integrating by part and removing $\alpha^{\prime\prime}$ and
$\beta^{\prime\prime}$, the effective counterpart of equations
(\ref{lone}) and (\ref{ltwo}) will have the form
\begin{equation}\label{eflone}
\mathcal{L}_N^{(1)
eff}=\frac{r^2}{G}\bigg\{\frac{V_1}{4}+\frac{(\alpha+3\beta)V_0}{4}+[14\beta+2\alpha-10\psi+12b^2(\alpha-\beta)]b^{\prime2}
$$$$-\frac{(b^2-1)\beta^{\prime}}{r}+\frac{(1-b^2)\alpha^{\prime}}{r}+8b(1-b^2)b^{\prime}\alpha^{\prime}+2b(5-4b^2)b^{\prime}\beta^{\prime}\bigg\}
\end{equation}
and
\begin{equation}\label{efltwo}
\mathcal{L}_N^{(2)eff}=\frac{r^2}{G}\bigg\{
\frac{V_2}{4}+\frac{(\alpha+3\beta)V_1}{4}+\bigg(16b^4-16b^2+\frac{1}{2}\bigg)\alpha^\prime\beta^\prime+(9-8b^2)(b^2-1)\beta^{\prime2}
$$$$-(b^2-1)b^{\prime}\psi^{\prime}
+
\frac{(1-b^2)\omega\psi^{\prime2}}{2}+\frac{[\psi+\alpha-\beta-1+b^2(\beta-3\alpha)]\alpha^\prime}{r}+\frac{\alpha^\prime\psi^\prime}{2}$$$$
+\frac{(\beta-\alpha-2\psi)(b^2-1)\beta^\prime}{r}+[4(\alpha+3\beta)(3\alpha+4\beta)+2\psi(5\alpha+\beta)$$$$
+12b^2(\alpha-\beta)(\alpha+3\beta-\psi)]b^{\prime2}+4[9\beta+3\psi-\alpha+2b^2(5\beta-3\alpha-\psi)]bb^{\prime}\alpha^{\prime}$$$$+\bigg(\frac{1}{2}-2b^2-
8b^4\bigg)\alpha^{\prime2}
+2[\alpha+59\beta-5\psi+4b^2(\psi-11\beta+\alpha)]bb^{\prime}\beta^{\prime}
\bigg\}.
\end{equation}
Without reducing the generality of the issue, we assume that the
interaction potential has the following form
\begin{equation}\label{pot}
GU(\phi,N^{\mu})\approx V_0(b)+\epsilon
V_1(\phi,N_\mu)+\epsilon^2V_2(\phi,N_\mu)+O(\epsilon^3)
 \end{equation}
Adding (\ref{bdeff}), (\ref{lzero}), (\ref{eflone}) and
(\ref{efltwo}) we can obtain total effective Lagrangian density of
the system as follows
\begin{equation}\label{teff}
G\mathcal{L}_{tot}^{eff}=L_{tot}^{(0)}+\epsilon
L_{tot}^{(1)}+\epsilon^2L_{tot}^{(2)}+O(\epsilon^3)
\end{equation}

where
    \begin{equation}\label{harm}
    L_{tot}^{(0)}=r^2\bigg(\frac{V_0}{4}-10b^{\prime2}\bigg), \end{equation}
\begin{equation}\label{order1}
L_{tot}^{(1)}=r^2\bigg\{\frac{V_1}{4}+\frac{(\alpha+3\beta)V_0}{4}+[14\beta+2\alpha-10\psi+12b^2(\alpha-\beta)]b^{\prime2}
$$$$-\frac{(b^2-1)\beta^{\prime}}{r}+\frac{(1-b^2)\alpha^{\prime}}{r}+8b(1-b^2)b^{\prime}\alpha^{\prime}+2b(5-4b^2)b^{\prime}\beta^{\prime}
 \bigg\}\end{equation}
 and
\begin{equation}\label{order2}L_{tot}^{(2)}=r^2\bigg\{
\frac{V_2}{4}+\frac{(\alpha+3\beta)V_1}{4}+\bigg(16b^4-16b^2-2\bigg)\alpha^\prime\beta^\prime-(8b^4-17b^2+12)\beta^{\prime2}
$$$$-(b^2-1)b^{\prime}\psi^{\prime}+
\frac{(1-2b^2)\omega\psi^{\prime2}}{4}+\frac{[\psi+\alpha-\beta-1+b^2(\beta-3\alpha)]\alpha^\prime}{r}$$$$
-\beta^\prime\psi^\prime+\frac{(\beta-\alpha-2\psi)(b^2-1)\beta^\prime}{r}+[4(\alpha+3\beta)(3\alpha+4\beta)+2\psi(5\alpha+\beta)$$$$
+12b^2(\alpha-\beta)(\alpha+3\beta-\psi)]b^{\prime2}+4[9\beta+3\psi-\alpha+2b^2(5\beta-3\alpha-\psi)]bb^{\prime}\alpha^{\prime}$$$$-
2b^2(1+4b^2)\alpha^{\prime2}
+2[\alpha+59\beta-5\psi+4b^2(\psi-11\beta+\alpha)]bb^{\prime}\beta^{\prime}
\bigg\}\end{equation}
 for
\begin{equation}\label{pot12}
V_1=0,~~~V_2=0.
\end{equation}
 The effective total lagrangian
(\ref{teff}) shows that in the weak field limit
 and slowly varying Brans-Dicke scalar field where
 $\epsilon(\omega)\to0$ and  $\omega\to\infty$ the  lagrangian density of the vector field is dominated
 instead of the Brans Dicke scalar field lagrangian density. In fact
for $\omega\to\infty$ the Brans-Dicke action (\ref{ac11}) reaches
to the Einstein-Hilbert counterpart which can not support the
galactic rotation curves alone. In the weak field limits
$(\epsilon\to0)$ one can see that the Brans Dicke Lagrangian
counterpart (\ref{bdeff}) vanishes and so the vector field
lagrangian density (\ref{harm}) is dominated to determine the
galactic rotation curves at $\epsilon=0.$ Now we obtain Euler
lagrange equations of the fields $b(r),~ \alpha(r),~\beta(r)$ and
$\psi(r)$ by varying the total Lagrangian density (\ref{teff}) and
solve them order by order as follows. Setting $\epsilon=0$ the
zero order term of the Euler-Lagrange equation reads
\begin{equation}\label{varp}
b^{\prime\prime}+\frac{2b^{\prime}}{r}+\frac{1}{80}\frac{\partial
V_0(b)}{\partial b}=0
\end{equation} which is independent of other fields and so can be solved alone. To solve the above equation we should
choose some suitable potentials for which solutions of the above
equation satisfy the time-like condition $q=\sqrt{b^2-1}$ (i.e.
$|b|>1)$ with boundary condition $q(\infty)=0.$ Thus we choose the
following ansatz for the potential $V_0(b).$
\begin{equation}\label{qeq}V_0(b)=80\mu^2b-40\mu^2 b^2\end{equation} in which $\mu$ can be described as mass of the vector field
$N_{\mu}.$ We will see that $\mu$ is related with a suitable
effective cosmological constant $\Lambda$ which means the Hubble
constant of the expansion of the universe. Also the above
potential makes the equation (\ref{varp}) as linear differential
equation. We should point that, nonlinear differential equations
in general form, have usually unstable solutions which may be
reach to chaos. Substituting the above potential into the equation
(\ref{varp}) one can obtain a particular convergent solution for
$b(r)$ as follows.
\begin{equation}\label{qeqh} b(r)=1+A\frac{e^{-\mu r}}{r}\end{equation}
where $A$ is integral constant which should be considered as
fitting parameter when we study its effect on the galactic
circular velocity. The equation (\ref{varp}) with potential
$(\ref{qeq})$ has other solution as $\frac{e^{\mu r}}{r}$ for
$\mu>0$ which we do not consider here because it diverges to
infinity at $r\to\infty.$ It is clear that only the decaying
exponentials should be a part of the physical solution where the
energy does not diverge to some infinite values and  as we should
retrieve the standard Newtonian gravitational potential
    at $r\to\infty$. Substituting (\ref{qeqh}), the equation $q(r)=\pm\sqrt{b^2-1}$ reads \begin{equation}\label{qsolv}
q(r)=\pm \sqrt{\frac{A^2e^{-2\mu r}}{r^2}+\frac{2Ae^{-\mu r}}{r}}
\end{equation}
which satisfies the boundary condition $q(\infty)=0.$ At large
distances $r\to\infty$ we see the vector field components approach
to $(b,q)\approx(1,0)$ and $(b^{\prime},q^{\prime})\approx(0,0).$
Substituting these asymptotic solutions into the effective
Lagrangian densities (\ref{harm}), (\ref{order1}) and
(\ref{order2}) we obtain asymptotic behavior of the total
effective Lagrangian density (\ref{teff}) as follows
\begin{equation}\label{asymplag}
\lim_{r\to\infty}G\mathcal{L}_{tot}^{eff}\approx10\mu^2
r^2[1+\epsilon(\alpha+\beta)]+\epsilon^2
r^2\bigg[3\beta^{\prime2}-10\alpha^{\prime2}$$$$
-\frac{\omega\psi^{\prime2}}{4}+\frac{(\psi-2\alpha-1)\alpha^\prime}{r}-2\alpha^\prime\beta^{\prime}-\psi^{\prime}\beta^\prime\bigg]
\end{equation}
Varying the Lagrangian density (\ref{asymplag}) with respect to
the fields $\psi(r), \beta(r)$ and $\alpha(r)$ we obtain their
Euler-Lagrange equations which up to second order terms
$O(\epsilon^2)$ become respectively as
\begin{equation}\label{psi}
2r\alpha^{\prime}+\omega(r^2\psi^\prime)^\prime\approx0
\end{equation}
\begin{equation}\label{beta}
10\mu^2
r^2+\epsilon[r^2(\psi+2\alpha-6\beta)^\prime]^\prime\approx0
\end{equation}
and
\begin{equation}\label{alpha}10\mu^2
r^2+\epsilon\{-2r\alpha^\prime+[2r^2(10\alpha+\beta)^\prime+(1+2\alpha-\psi)r]^\prime\}\approx0.
\end{equation}
To solve the above equations we first substitute
$2r\alpha^{\prime}$ from (\ref{psi}) into the equation
(\ref{alpha}) and integrate it to obtain
\begin{equation}\label{newalpha} \frac{10}{3}\mu^2 r^2+\frac{C_1}{r}+\epsilon[r(\omega\psi+20\alpha+2\beta)^\prime+1-\psi+2\alpha]\approx0 \end{equation}
in which $C_1$ is integral constant. The equation (\ref{beta})
 can be integrated alone to obtain
\begin{equation}\label{newbeta} \psi+2\alpha-6\beta=C_3+\frac{C_2}{r}-\frac{5}{3}\mu^2 r^2.\end{equation} Now we substitute $\beta$ from the above
 equation and $r\alpha^{\prime}$ from the equation (\ref{psi}) into the equation (\ref{newalpha}) to obtain
\begin{equation} \label{psialfa}\frac{10}{3}\bigg(1+\frac{\epsilon}{3}\bigg)\mu^2 r^2+\frac{(3C_1+\epsilon C_2)}{3r}+
\epsilon+\frac{\epsilon(1-59\omega)}{3}r\psi^\prime-\frac{31\omega\epsilon}{3}r^2\psi^{\prime\prime}-\epsilon\psi+2\epsilon\alpha\approx0
\end{equation}
in which $C_{2,3}$ are integral constants. Substituting
$\alpha^{\prime}$ from the equation (\ref{alpha}) into the
derivative of the above equation we obtain a linear differential
equation for the Brans Dicke scalar field $\psi(r)$ as follows.
\begin{equation}\label{psiequation}
\psi^{\prime\prime\prime}+\bigg(4-\frac{1}{31\omega}\bigg)\frac{\psi^{\prime\prime}}{r}+\bigg(\frac{65}{31}+\frac{2}{31
\omega}\bigg)\frac{\psi^\prime}{r^2}+\frac{(3C_1+\epsilon
C_2)}{31\omega\epsilon}\frac{1}{r^4}-\frac{20(3+\epsilon)}{93\epsilon\omega}\frac{\mu^2}{r}\approx0.
\end{equation}
One can solve explicitly the equations (\ref{psiequation}),
(\ref{psialfa}) and (\ref{newbeta}) synchronously to obtain exact
solutions for the fields $\psi(r),$ $\alpha(r)$ and $\beta(r)$
respectively as follows.
\begin{equation}\label{betaapp}
\epsilon\psi(r)=\frac{(3C_1+\epsilon
C_2)}{(4+3\omega)}\frac{1}{r}+\frac{10(3+\epsilon)}{(1+189\omega)}\frac{\mu^2
r^2}{3}+\epsilon C_4\frac{r^{f_-(\omega)}}{f_-(\omega)}+\epsilon
C_5\frac{r^{f_+(\omega)}}{f_+(\omega)}+\epsilon C_6\end{equation}
where $C_{4,5,6}$ are integral constants and we defined
\begin{equation}\label{diff}f_{\pm}(\omega)=-\frac{1}{2}+\frac{1}{62\omega}\pm\sqrt{589-\frac{434}{\omega}+\frac{1}{\omega^2}}\end{equation}
and \begin{equation}\label{alphaapp}\epsilon\alpha(r)=-K_1 \mu^2
r^2+\frac{K_2}{r}+\epsilon
C_4\frac{r^{f_-(\omega)}}{g_-(\omega)}+\epsilon
C_5\frac{r^{f_+(\omega)}}{g_+(\omega)}+\frac{\epsilon (C_6-1)}{2}
\end{equation}
where we defined
\begin{equation}\label{K1}K_1=
\frac{5}{9}\frac{(378\epsilon^2\omega+2\epsilon^2+954\epsilon\omega+5\epsilon-540\omega-3)}{(1+189\omega)},~~~K_2=\frac{(1-2\epsilon)(3C_1+\epsilon
C_2)}{6}\end{equation} with
\begin{equation}\label{g}g_{\pm}(\omega)=\frac{f_{\pm}(\omega)}{31\omega f^2_{\pm}(\omega)+(28\omega-1)f_{\pm}(\omega)+3}\end{equation}
\begin{equation}\label{alphasol}
\epsilon \beta(r)=K_3\mu^2
r^2+\frac{\epsilon(2C_6-C_3-1)}{6}+\frac{K_4}{r}+\epsilon
C_4\frac{r^{f_-(\omega)}}{h_-(\omega)}+\epsilon
C_5\frac{r^{f_+(\omega)}}{h_+(\omega)}\end{equation} in which we
defined
\begin{equation}\label{K_3}K_3=\frac{5}{9}\bigg(\frac{3+\epsilon}{1+189\omega}\bigg)-\frac{K_1}{3}+\frac{5}{18}\end{equation}
\begin{equation}\label{K_4}
K_4=\frac{3C_1[7-8\epsilon+3\omega(1-2\epsilon)]-\epsilon
C_2[209+8\epsilon+3\omega(2\epsilon+53)]}{18(4+3\omega)}\end{equation}
and
\begin{equation} \label{hplus}\frac{6}{h_{\pm}(\omega)}=\frac{1}{f_{\pm}(\omega)}+\frac{2}{g_{\pm}(\omega)}.\end{equation}
 Substituting the above
solutions into the relations (\ref{sera}), (\ref{serb}) and
(\ref{serp}) we obtain  respectively
\begin{equation}\label{metalpha}e^{2\epsilon\alpha(r)}\sim1+\epsilon
(C_6-1)-2K_1 \mu^2 r^2+\frac{2K_2}{r}+2\epsilon
C_4\frac{r^{f_-(\omega)}}{g_-(\omega)}+2\epsilon
C_5\frac{r^{f_+(\omega)}}{g_+(\omega)}\end{equation}
\begin{equation}\label{metbeta}e^{2\epsilon\beta(r)}\sim
1+\frac{\epsilon(2C_6-C_3-1)}{3}+2K_3\mu^2
r^2+\frac{2K_4}{r}+2\epsilon
C_4\frac{r^{f_-(\omega)}}{h_-(\omega)}+2\epsilon
C_5\frac{r^{f_+(\omega)}}{h_+(\omega)}
\end{equation} and
\begin{equation}\label{phisoll}G\phi(r)\sim1+\epsilon C_6+\frac{(3C_1+\epsilon
C_2)}{(4+3\omega)}\frac{1}{r}+\frac{10(3+\epsilon)}{(1+189\omega)}\frac{\mu^2
r^2}{3}+\epsilon C_4\frac{r^{f_-(\omega)}}{f_-(\omega)}+\epsilon
C_5\frac{r^{f_+(\omega)}}{f_+(\omega)} \end{equation} Now we
should fix the integral constants of the above solutions by
regarding physical boundary conditions.\\ Appearance of the square
term in the above metric potentials as $\mu^2 r^2$ remember us the
 de Sitter universe at large scale structure of the metric
solution where $\mu^2$ behaves as an effective cosmological
constant $\Lambda>0$. This means that our solutions treat as de
Sitter metric at large distances $\mu r>>1$ while at small
distances $\mu r<<1$ the inverse distance factor $\frac{1}{r}$ is
dominated instead of the term $\mu^2 r^2$ which can be related to
the Schwarzschild counterpart of the metric in absence of the
terms $C_{4,5}$. We know that the Schwarzschild de Sitter black
hole metric with $\Lambda>0$ is $ds^2=-(1-2GM/r-\Lambda
r^2/3)dt^2+dr^2/(1-2GM/r-\Lambda r^2/3)+r^2
d\theta^2+r^2\sin^2\theta d\varphi^2$ which is obtained from
$G_{\mu\nu}+\Lambda g_{\mu\nu}=0$ and  asymptotically reduces to
the vacuum de Sitter space at large distances  $r>>2GM.$ In weak
field limit it reads $ds^2=-(1-2GM/r-\Lambda
r^2/3)dt^2+(1+2GM/r+\Lambda r^2/3)dr^2+r^2
d\theta^2+r^2\sin^2\theta d\varphi^2$ which can be compared with
solutions (\ref{metalpha}) and (\ref{metbeta}) by setting
\begin{equation} \label{cosmo}2K_1\mu^2=\frac{\Lambda}{3}=2K_3\mu^2,~~~2K_4=2GM=-2K_2,~~~C_6=C_3=1\end{equation}
where $M$ is total mass of the central black hole of the
Schwarzschild de Sitter space time. In the standard $\Lambda CDM$
cosmological model the unknown exotic dark matter/energy is
dominated to support the cosmic inflation and it behaves as an
effective cosmological constant.
 Substituting (\ref{K1}), (\ref{K_3}), (\ref{K_4})
into the condition (\ref{cosmo}) the equation $K_3=K_1$ reads
\begin{equation}\label{epsilon}\epsilon_{\pm}(\omega)=\frac{-(17+3816\omega)\pm\sqrt{29340144\omega^2+344016\omega+1009}}{16(1+189\omega)}\end{equation}
and the equation $K_2+K_4=0$ reduces to the following condition.
\begin{equation}\label{C12}\frac{C_1}{C_2}=\frac{\epsilon[197+150\omega+8\epsilon(4+3\omega)]}{3[19+12\omega-8\epsilon(4+3\omega)]}=S(\omega)
\end{equation} in which \begin{equation}\label{C2}C_2=\frac{-6GM}{(1-2\epsilon)(3S(\omega)+\epsilon)}.\end{equation}
Substituting (\ref{cosmo}) into the solutions (\ref{metalpha}),
(\ref{metbeta}) and (\ref{phisoll}) we will have
\begin{equation}\label{metalpha1}e^{2\epsilon\alpha(r)}\sim1-\frac{\Lambda}{3} r^2-\frac{2GM}{r}+2\epsilon
C_4\frac{r^{f_-(\omega)}}{g_-(\omega)}+2\epsilon
C_5\frac{r^{f_+(\omega)}}{g_+(\omega)}
\end{equation}
\begin{equation}\label{metbeta2}e^{2\epsilon\beta(r)}\sim
1+\frac{\Lambda }{3}r^2+\frac{2GM}{r}+2\epsilon
C_4\frac{r^{f_-(\omega)}}{h_-(\omega)}+2\epsilon
C_5\frac{r^{f_+(\omega)}}{h_+(\omega)}
\end{equation} and
\begin{equation}\label{phisoll22}G\phi(r)\sim1+\epsilon-\frac{6}{(1-2\epsilon)(4+3\omega)}\frac{GM}{r}+J(\omega)\Lambda r^2
+\epsilon C_4\frac{r^{f_-(\omega)}}{f_-(\omega)}+\epsilon
C_5\frac{r^{f_+(\omega)}}{f_+(\omega)} \end{equation} in which we
defined
\begin{equation}\label{J}J(\omega)=\frac{3+\epsilon}{378\epsilon^2\omega+2\epsilon^2+954\epsilon\omega+5\epsilon-540\omega-3}.\end{equation}
 Now we should say about physical situations of the integral constants $C_{4,5}.$ We can show that they are related to a radial length
  $r_{qf}$ scale
 which determines quasi flat regions of the space time where \begin{equation}\label{qf}
 e^{2\epsilon\alpha(r_{qf})}=1=e^{2\epsilon\beta(r_{qf})}\end{equation} for which the equations (\ref{metalpha1}) and (\ref{metbeta2}) read
\begin{equation}C_4=\frac{\bigg(\frac{1}{h_+(\omega)}+\frac{1}{g_+(\omega)}\bigg)\bigg(\frac{\Lambda r_{qf}^2}{3}+
\frac{2GM}{r_{qf}}\bigg)}{2\epsilon
r_{qf}^{f_-(\omega)}\bigg(\frac{1}{g_-(\omega)h_+(\omega)}
-\frac{1}{g_+(\omega)h_-(\omega)}\bigg)}\end{equation} and
\begin{equation}C_5=-\frac{\bigg(\frac{1}{h_-(\omega)}+\frac{1}{g_-(\omega)}\bigg)\bigg(\frac{\Lambda r_{qf}^2}{3}+
\frac{2GM}{r_{qf}}\bigg)}{2\epsilon
r_{qf}^{f_+(\omega)}\bigg(\frac{1}{g_-(\omega)h_+(\omega)}
-\frac{1}{g_+(\omega)h_-(\omega)}\bigg)}.\end{equation} We know
 from the equation (\ref{om}) which the solution (\ref{phisoll22}) should not diverge to infinity at $\omega\to\infty$.
 In fact for quasi flat regions the solution (\ref{phisoll22}) should reach to the boundary condition (\ref{om}) for which we should set
 \begin{equation}\label{lim}\lim_{\omega\to\infty}\epsilon-\frac{6}{(1-2\epsilon)(4+3\omega)}\frac{GM}{r_{qf}}+J(\omega)\Lambda r_{qf}^2
+\epsilon C_4\frac{r_{qf}^{f_-(\omega)}}{f_-(\omega)}+\epsilon
C_5\frac{r_{qf}^{f_+(\omega)}}{f_+(\omega)}\to0
\end{equation} It is easy to check for large $\omega$ we will have asymptotically
 \begin{equation} f_{-}(\omega)\approx-24.769+\frac{8.9573}{\omega},~~~f_+(\omega)\approx23.769-\frac{8.9251}{\omega}\end{equation}
 \begin{equation}g_-(\omega)\approx-\frac{0.0013517}{\omega},~~~g_+(\omega)\approx\frac{0.0013074}{\omega}\end{equation}
 and \begin{equation}h_-(\omega)\approx-\frac{0.0040552}{\omega},~~~~~h_+(\omega)\approx\frac{0.0039223}{\omega}\end{equation}
 \begin{equation}\label{epsilonm}\epsilon_-(\omega)\approx-3.0531+\frac{0.0000314}{\omega},~~~\epsilon_+(\omega)\approx0.5293+\frac{0.0020788}{\omega}
 \end{equation}
\begin{equation}\frac{C_1(\omega)}{GM}\approx0.15+\frac{0.005}{\omega},~~~\frac{C_2(\omega)}{GM}-46.15-\frac{1.02}{\omega}\end{equation}
 \begin{equation}J_-(\omega)\approx-\frac{0.00075}{\omega},~~~J_+(\omega)\approx\frac{0.049849}{\omega}\end{equation}
 \begin{equation}C_4^+=\frac{6112.788719}{\omega}\bigg(\frac{\frac{\Lambda}{3}r_{qf}^2+\frac{2GM}{r_{qf}}}{r^{-24.769+\frac{8.9573}{\omega}}_{qf}}
 \bigg),~~~~\frac{C^-_4}{C^+_4}\approx-0.1734\end{equation}
 \begin{equation}C_5^+=-\frac{5912.452201}{\omega}\bigg(\frac{\frac{\Lambda}{3}r_{qf}^2+\frac{2GM}{r_{qf}}}{r^{23.769-\frac{8.9251}{\omega}}_{qf}}
 \bigg),~~~\frac{C_5^-}{C_5^+}\approx-0.1734\end{equation} and \begin{equation}\label{mu}\mu^2_-\approx\frac{\Lambda}{3}\bigg[2.4012+\frac{0.001356}{\omega}
 \bigg],~~~\mu^2_+\approx\frac{\Lambda}{3}\bigg[2.4012-\frac{0.08973}{\omega}\bigg].\end{equation}
 The above calculations show that for $\omega\to\infty$ we will have
 $J(\infty)=0=C_{4,5}(\infty)$ and so the equation (\ref{lim}) leads to small $\epsilon_+(\infty)\approx+0.5$ instead a zero value.
 This means the choice $\epsilon_+$ is physical but not $\epsilon_-.$ In fact, if we substitute $\epsilon_-(\infty)\approx-3$ into the equation
 (\ref{lim}) then we can result as $\lim_{\omega\to\infty} G\phi\to-2$ which is not a physical boundary condition ($\phi$ is a positive valued field).
However for a large $\omega$ we substitute the above approximated
equations together with
$\epsilon_+\approx0.5293+\frac{0.0020788}{\omega}$ into the
equation (\ref{lim}) to obtain an equation for $r_{qf}$ as
follows.
\begin{equation}\label{rqf}0.5293\omega-\frac{490.4471GM}{r_{qf}}-29.1413\frac{\Lambda
r^2_{qf}}{3}\approx0\end{equation} Regarding $\vartheta$ as a free
parameter one can obtain the following identities from
(\ref{rqf}).
\begin{equation}\label{vartheta}\frac{2GM}{r_{qf}}=0.0022\omega\cos^2\vartheta,~~~\frac{\Lambda
r_{qf}^2}{3}=0.0020\omega\sin^2\vartheta.
\end{equation} It is remarkable that the left side of the above equation shows
that for large $\omega$ the quasi flat region $r_{qf}$ should
reach to the Schwarzschild radius $2GM$ of the central black hole
of the galaxy under consideration while the right side of them
shows that by raising $\omega$ the quasi flat region $r_{qf}$
reach to the cosmological event horizon of the de Sitter vacuum
space $\sqrt{\frac{3}{\Lambda}}\approx1.833\times10^9 (light~
year)=0.5623\times10^9(Parsec).$ In other words for a fixed
$\omega$ the left side equation shows that by rasing $r_{qf}$ then
$\vartheta\to\frac{\pi}{2}.$ These helps us to obtain an
approximated solution for $r_{qf}$ by setting
$\vartheta=\frac{\pi}{2}$ in which we have
\begin{equation}\label{rqf2}r_{qf}\approx0.23343\sqrt{\frac{\omega}{\Lambda}}=7.5785\times10^4\sqrt{\omega}(Kpc).\end{equation}
At last we are in a position to write metric solutions for large
$\omega$ as follows.
\begin{equation}\label{metalphaplus}e^{2\epsilon\alpha_+(r)}\sim1-\frac{\Lambda r^2}{3}-\frac{2GM}{r}
-1.6\times10^6\Lambda
r_{qf}^2\bigg[\bigg(\frac{r_{qf}}{r}\bigg)^{25}+\bigg(\frac{r}{r_{qf}}\bigg)^{24}\bigg]\end{equation}
and \begin{equation}\label{metbetaplus}e^{2\epsilon
\beta_+(r)}\sim1+\frac{\Lambda r^2}{3}+\frac{2GM}{r}
-0.534\times10^6\Lambda
r^2_{qf}\bigg[\bigg(\frac{r_{qf}}{r}\bigg)^{25}+\bigg(\frac{r}{r_{qf}}\bigg)^{24}\bigg]\end{equation}

 In the following section we  obtain equation of motion of a test
particle orbiting around  galaxy in weak limit of the
gravitational field.
\section{Modified equations of motion}
For galactic scales studies one can consider a test particle (a star in our consideration) as a
tracer of gravitational field.
At first approximation we often assume the galaxies are
spherically symmetric objects. When we investigate rotation curves
at significantly larger radii than the central region of the
galaxy, it is justifiable to use a point-like source approximation
for any physical model that does not involve an extended dark
matter halo. These approximations are surprisingly useful because
of the often very large uncertainty in the baryonic mass-to-light ratio(M/L) of many galaxies.
We assume a test star with mass parameter $m$ orbiting around a
central body and interact with the 4-vector field $N^{\mu}$ for
which one can write its action functional as follows
\begin{equation}\label{tpac}
I_{TP}=-m\int d\tau\sqrt{g_{\mu\nu}V^\mu V^\nu}+\lambda\int d\tau
N_\mu V^{\mu}
\end{equation}
where $N_\mu=(b(r),q(r),0,0)$ given by (\ref{vec}) is 4-velocity
of the preferred reference frame, $\tau$ is the proper time along
the world line of the test particle with the dimension of length.
$\lambda$ denotes to the interaction coupling constant between the
vector field $N^{\mu}$ and the test particle with four velocity
$V^{\mu}=\frac{dx^\mu}{d\tau}$. Varying (\ref{tpac}) with respect
to the coordinates $x^{\mu}$ one can obtain Euler Lagrange
equations of the test particle such that
\begin{equation} \label{eqsmotion}
 \frac{d^2x^\mu}{d\tau^2}
+\Gamma^\mu_{\alpha\beta}\frac{dx^\alpha}{d\tau}\frac{dx^\beta}{d\tau}
=\frac{\lambda}{m}{F^\mu}_\nu\frac{dx^\nu}{d\tau}.
\end{equation}
For the line element (\ref{meta}), the equation of motion
(\ref{eqsmotion}) in weak field approximation for small $\epsilon$
read
    \begin{equation}\label{gt2}
    \frac{d^2t}{d\tau^2}+\frac{2\epsilon\alpha^{\prime}}{1+2\epsilon \alpha}
    \biggl(\frac{dt}{d\tau}\biggr)\biggl(\frac{dr}{d\tau}\biggr)  \cong\frac{2\lambda(1-2\epsilon\alpha(r))}{m}b^{\prime}(r)\biggl(\frac{dr}{d\tau}\biggr),
    \end{equation}
    \begin{equation}\label{gr2}
    \frac{d^2r}{d\tau^2}+\frac{\epsilon\beta^{\prime}}{1+2 \epsilon \beta}\bigg(\frac{dr}{d\tau}\bigg)^2+\frac{\epsilon\alpha^{\prime}}{1+
    2 \epsilon \beta}
    \bigg(\frac{dt}{ d\tau}\bigg)^2-\frac{(r+\epsilon
    r^2\beta^{\prime}+2\epsilon r\beta)}{r+2 \epsilon \beta}\bigg(\frac{d\theta}{d\tau}\bigg)^2 $$$$
-r\sin^2\theta\frac{(1+\epsilon \beta' r+2 \epsilon \beta)}{1+2
\epsilon \beta}\bigg(\frac{d\varphi}{d\tau}\bigg)^2
    =\frac{2\lambda(1-2\epsilon\beta(r))}{m}
b^{\prime}\bigg(\frac{dt}{d\tau}\bigg)
    \end{equation}
    \begin{equation}\label{gte22}
    \frac{d^2\theta}{d\tau^2}+\frac{2(1+\epsilon
        r\beta^{\prime}+2\epsilon \beta)}{r+2\epsilon \beta r}\biggl(\frac{d\theta}{d\tau}
    \biggr)\biggl(\frac{dr}{d\tau}\biggr)-\sin\theta\cos\theta\biggl(\frac{d\varphi}{d\tau}\biggr)^2
    =0,
    \end{equation}
    and
    \begin{equation}\label{gph22} \frac{d^2\varphi}{d\tau^2}+\frac{2(1+\epsilon
        r\beta^{\prime}+2 \epsilon \beta)}{r+2\epsilon \beta r}\biggl(\frac{d\varphi}{d\tau}\biggr)\biggl(\frac{dr}{d\tau}\biggr)
    +\cot\theta\biggl(\frac{d\varphi}{d\tau}\biggr)\biggl(\frac{d\theta}{d\tau}\biggr)=0.
    \end{equation}
Because of spherically symmetric property of the metric equation,
we choose $\theta=\frac{\pi}{2}$ for planar orbit of the test
particle which satisfies trivially equation (\ref{gte22}).
Therefore equation (\ref{gph22}) leads to a conserved angular
momentum of the test particle as follows
\begin{equation}\label{ang}
\frac{d\varphi}{d\tau}=\frac{L}{m(1+2\epsilon \beta)}\frac{1}{r^2}
\end{equation}
in which $L$
 is constant angular momentum of the test particle. In the weak field limit, slow motions of the test particle and
 for circular orbits we can substitute the approximations
 $\frac{dt}{d\tau}\approx1$ and $\frac{dr}{d\tau}\approx0$ into the geodesic equations (\ref{gt2}) and (\ref{gr2}). Equation (\ref{gt2})
 vanishes trivially
 while equation (\ref{gr2}) leads to the acceleration equation of the test particle, which up to
second order term $O(\epsilon^2)$ is
 \begin{equation}\label{acc}
\frac{d^2r}{dt^2}\approx-\epsilon\alpha^{\prime}+\frac{L^2[1-4\epsilon\beta+\epsilon
r\beta^{\prime}]}{m^2r^3} +\frac{2\lambda
b^{\prime}[1-2\epsilon\beta]}{m}. \end{equation}
 To obtain circular stable orbits
of test particle (the test star) we should set
$\frac{dr^2}{dt^2}=0$
 and $L=mrv(r)$ in the above equation where $v(r)$  is circular velocity of the test particle moving on a circular orbit with constant radius $r.$
Regarding the latter conditions the equation (\ref{acc}) reads
\begin{equation}\label{velocity}v^2(r)=\frac{-\frac{2\lambda rb^{\prime}}{m}+\epsilon(r\alpha^{\prime}+\frac{4\lambda r \beta
b^{\prime}}{m})}{1+\epsilon(r\beta^{\prime}-4\beta)}.
\end{equation}
Substituting (\ref{qeqh}),  (\ref{metalphaplus}) and
(\ref{metbetaplus}) into the above formula we obtain galactic
circular velocity as follows.
\begin{equation}\label{velocityr}v(r)\approx \sqrt{\frac{GM}{r}}\sqrt{\frac{P(r)}{Q(r)}}\end{equation} where we
defined \begin{equation}\label{PP}
P(r)=1+2\tan^2\vartheta\bigg[60\bigg(\frac{r_{qf}}{r}\bigg)^{24}-57\bigg(\frac{r}{r_{qf}}\bigg)^{25}-\bigg(\frac{r}{r_{qf}}\bigg)^3\bigg]$$$$
+\sigma
\bigg(1+\frac{r}{r_{qf}}\sqrt{0.002\omega}\sin\vartheta\bigg)e^{-\frac{r}{r_{qf}}\sqrt{0.002\omega}\sin\vartheta}$$$$\times\bigg\{1-0.002\omega\cos^2\vartheta\bigg(\frac{r_{qf}}{r}\bigg)+0.002\omega\sin^2\vartheta
\bigg[\frac{3}{2}\bigg(\frac{r_{qf}}{r}\bigg)^{25}+\frac{3}{2}\bigg(\frac{r}{r_{qf}}\bigg)^{24}-\bigg(\frac{r}{r_{qf}}\bigg)^2\bigg]\bigg\}\end{equation}
and
\begin{equation}\label{QQ} Q(r)=1-0.005\omega\cos^2\vartheta\bigg(\frac{r_{qf}}{r}\bigg)+0.002\omega\sin^2\vartheta
\bigg[24\bigg(\frac{r_{qf}}{r}\bigg)^{25}-18\bigg(\frac{r}{r_{qf}}\bigg)^{24}-\bigg(\frac{r}{r_{qf}}\bigg)^2\bigg]\end{equation}
 where  we used (\ref{vartheta}) and \begin{equation}\label{def}\sigma=\frac{2A\lambda}{GmM}.\end{equation} It is easy to check  that
 the above formula for the circular velocity reaches to the Newtonian regime $v_N=\sqrt{\frac{GM}{r}}$ at the quasi flat regions $r\approx r_{qf},$
 if we set \begin{equation}P(r_{qf})=Q(r_{qf})\end{equation} which reads \begin{equation}\label{angel}
 3\sin^4\vartheta+4\bigg(\frac{200}{\omega}-1
 \bigg)\sin^2\vartheta+1\approx0\end{equation} for large $\omega.$ The equation (\ref{angel}) has two different solution as
 \begin{equation}\label{vartheta1}\sin^2\vartheta_1\approx1-\frac{200}{\omega}+\cdots
 \end{equation} and \begin{equation}\label{vartheta2}\sin^2\vartheta_2\approx\frac{1}{3}+\frac{200}{3\omega}+\cdots.\end{equation} $\vartheta_1$
 takes $\frac{\pi}{2}$ for $\omega\to\infty$ and so it is not a suitable physical case
 because $\tan \vartheta$ given in the equation (\ref{PP}) diverges to a infinite value. Thus we will continue the work by using
 (\ref{vartheta2}) to calculate the galactic circular velocity (\ref{velocityr})
 which for large $\omega$ we choose $\sin^2\vartheta_2\sim\frac{1}{3}$ where (\ref{PP}) and (\ref{QQ}) read respectively\begin{equation}\label{PPP}
P(r)=1+60\bigg(\frac{r_{qf}}{r}\bigg)^{24}-57\bigg(\frac{r}{r_{qf}}\bigg)^{25}-\bigg(\frac{r}{r_{qf}}\bigg)^3$$$$
+\sigma
\bigg(1+\frac{r}{r_{qf}}\sqrt{\frac{\omega}{1500}}\bigg)e^{-\frac{r}{r_{qf}}\sqrt{\frac{\omega}{1500}}}$$$$\times\bigg\{1+\frac{\omega}{1500}
\bigg[\frac{3}{2}\bigg(\frac{r_{qf}}{r}\bigg)^{25}+\frac{3}{2}\bigg(\frac{r}{r_{qf}}\bigg)^{24}-\bigg(\frac{r}{r_{qf}}\bigg)^2-2\bigg(\frac{r_{qf}}{r}\bigg)\bigg]\bigg\}\end{equation}
and
\begin{equation}\label{QQQ}
Q(r)=1+\frac{\omega}{1500}
\bigg[24\bigg(\frac{r_{qf}}{r}\bigg)^{25}-18\bigg(\frac{r}{r_{qf}}\bigg)^{24}-\bigg(\frac{r}{r_{qf}}\bigg)^2-5\bigg(\frac{r_{qf}}{r}\bigg)\bigg]
.\end{equation}  It is easy to check
\begin{equation}\label{rlarge}\lim_{r>>r_{qf}}\frac{P(r)}{Q(r)}\simeq\bigg(\frac{4750}{\omega}\bigg)\bigg(\frac{r}{r_{qf}}\bigg)\end{equation} and \begin{equation}\label{rsmall}\lim_{r<<r_{qf}}
\frac{P(r)}{Q(r)}\simeq\frac{\sigma}{16}.\end{equation}
  The above galactic circular velocity is for a point like source
with total mass $M.$ Let us now extend the equation
(\ref{velocityr}) for a mass distribution of spherically symmetric
galaxies where the mass function
\begin{equation}\label{mr} M(r)=4\pi\int_0^r\rho(\eta)\eta^2d\eta
\end{equation}
is the total amount of ordinary visible matter within a sphere of
radius $r$ and $\rho(\eta)$ is the density of visible matter of
the spherical galaxy contains an inner core at radius $r=r_c$.
 There are
different models which are proposed for mass distribution of
visible galaxies.
 According to the model presented in ref. \cite{BM} we consider here a
simple power-law mass distribution function as
\begin{equation}\label{masdis}
M(r)=M\bigg(\frac{r}{r_c+r}\bigg)^{3\varsigma}
\end{equation}
where $\varsigma$ is a constant parameter. The values for
$\varsigma$ are  $1$ and $2$ for HSB (high surface brightness)
 and LSB (low surface brightness) $\&$ Dwarf galaxies
respectively. This difference is due to the fact that rotation
curves
    of LSB and dwarf galaxies rise more slowly than those of HSB
    galaxies \cite{Blok}. $r_c$ is radius of the galaxy core. Inside the core $r<r_c$ there is a constant mass density for HSB
    galaxies. For LSB galaxies it is a raising function as $\rho(r)\approx (r/r_c)^3$
inside the core $r<r_c$ \cite{BM}. Well outside the core radius,
where $r >> r_c$, equation (\ref{masdis}) implies that
    \begin{equation}
    \underset{r>>r_c}{\text{lim}}M(r)=M
    \end{equation} in which $M$ is total mass of the galaxy under
    consideration. Numerical values for
$r_c$ are given in the table 1 for a set of 12 observed galaxies.
They are obtained by fitting the observational data and our
solutions for circular velocity of the galaxies called in the
table 1. At last we substitute (\ref{masdis}) into the circular
velocity given by (\ref{velocityr}) to obtain circular velocity
for a mass distributed galaxy as follows.
\begin{equation}\label{velocitydis}v(r)=\sqrt{\frac{GM}{r}}\bigg(\frac{r}{r_c+r}\bigg)^{\frac{3\varsigma}{2}}\sqrt{\frac{P(r)}{Q(r)}}\end{equation}
where $P(r)$ and $Q(r)$ should be inserted from (\ref{PPP}) and
(\ref{QQQ}). One can show that the Newton`s gravity coupling
constant $G\simeq6.67\times10^{-11} m^3/kg.s^2$ can be rewritten
 as $G=4.3\times10^{-6} \big(\frac{kpc}{M_{sun}}\big)\big(\frac{km}{s}\big)^2.$ In the latter case the galactic circular velocity can be described in
  units $km/s$ if we choose radial distance $r$ in units $kpc$ instead of the $metter$ and the
 galactic mass $M$ is given as relative mass with respect to the mass of sun as $(M/M_{sun})\times10^{10}.$
  Because $v=\sqrt{\frac{GM}{r}}\to\big(\frac{km}{s}\big)\sqrt{4.3\times10^{-6}\big(\frac{kpc}{r}\big)\big(\frac{M}{M_{sun}}\big)}.$
   To plot numerical diagram
  of the circular velocity (\ref{velocitydis}) we substitute radial distances in units $kpc$, and $G=4.3\times10^{-6}$ and $M$ as
  $(M/M_{sun}\times 10^{10})$ (see table 1). In the latter case the velocity is described in $(km/s)$ units and $r$ in units $kpc.$
  We plot numerical diagrams for the
Brans Dicke parameter as $\omega=40000$ (see refs.
\cite{CW1,CW2,EG,RD}). Other parameters given in the above formula
called as $r_{qf}, \sigma, r_c$ should be considered as fitting
parameters when we set result of our theoretical model with
observational data given in the table 1. We obtained numerical
values for the fitting parameters $(r_{qf}, \sigma, r_c)$ and
collect them into the table 1 for different galaxies which we used
to study. circular velocity diagrams for 12 observed galaxies are
given in the figure 1. There are three charts for each galaxy.
That is, the experimental data (black dots) and the theoretical
results (red dots) of our model and the Newtonian limit (blue
dots) of velocity are plotted in terms of distance from the center
of the galaxy. The galaxy mass is given from \cite{RH,Sobuti}.
There is two components for each galaxy mass which one of them is
related to the neutral Hydrogen gases $\mu_{HI}$ surrounds the
galaxy and the other is the central core star-like mass
$\mu_{star}$ and so we should use the galactic relative total mass
as $M=\mu_{HI}+\mu_{star}$ (see the table 1)  for each galaxy
rotation curve in the equation (\ref{velocitydis}).
\section{ Spiral galaxy rotation curves
and observational data}
    \subsection*{Sample selection}
We employ new database of SPARC (Spitzer Photometry and Accurate
Rotation Curves) \cite{SPARC} to select observed data of circular
velocity of a sample of 12 observed galaxies. They can be seen
with $black~dotes$ in figure 1. In fact SPARC is
  sample  of 175  nearby
galaxies with new surface photometry at 3.6 $\mu$m and
high-quality rotation curves from previous studies about the
 atomic hydrogen HI which is  one  of  the  best  kinematical  tracer  of  the  gravitational  potential  in  nearby
galaxies \cite{SPARC1}. In fact it is  representing all
rotationally supported morphological types of galaxies. To
minimize the star-halo degeneracy, the best approach is to use
near-infrared surface photometry (K-band or 3.6 $\mu$m), which
provides the closest proxy to the stellar mass (see \cite{SPARC1}
and references therein). In short the SPARC is the largest galaxy
sample to date with spatially resolved data on the distribution of
both stars and gas.  In order
    to test our model, we have therefore considered a sample of 12
    galaxies with well-measured rotation curves extracted
    from Ref.\cite{RH} and presented in table 1.

\subsection*{Model fit to rotation curves of galaxies}
 To investigate the rotation curves of galaxies
within the framework of JBD-SVT gravity, we suppose that there is
no actual dark matter, therefore, such a galaxy essentially
consists of baryonic matter containing stars and interstellar gas.
Hence the observed circular speeds and the Newtonian ones are
derived only from the observed mass of the stellar objects and HI
(Neutral Hydrogen) components of galaxies. Moreover we ignore dust
in our analysis, since the mass of the dust is at most a few
percent of the mass of the
interstellar matter.\\
Armed with equation (\ref{velocitydis}) we are in a position to
plots rotation curves for a
    sample of 12 spiral galaxies after determining the fitting parameters
   $\sigma, r_{c}, r_{qf}$ (see table 1).  To do so we use the `Nonlinear Model Fit` (NMF) function
    in Wolfram Mathematica software .
We observe very good agreement between the observed data points
and the fitted curve which are shown in figure 1.
 However, we must use
 a more sophisticated analysis for
    the general case where we do not have spherical symmetry, which is left for our next work. In fact the latter considerations bring some
    higher precision on the results.
     This is done for some spiral galaxies in ref. \cite{MRah} but for MOG gravity model.
It is worth to mention that for a broad range of even well
researched galaxies there
  is no consensus on galaxy mass among various sources. Since the
galaxy mass is not directly measured hence  their estimations are
based on various galactic models. So it is not surprising such a
broad range of estimations on the mass and core radius of
galaxies. However references of mass estimations
which we have used in this work, are mentioned in
 table 1 and extracted from refs. \cite{RH,Sobuti}.
    \begin{table}
        \centering
        \caption{Fitting parameters  $(\sigma, r_c, r_{qf})$  for 12 spiral galaxies. Their observed data which
        are used from \cite{RH,Sobuti} are the relative stellar baryonic mass $\mu_{stellar}=\frac{M_{stellar}}{M_{sun}}
        \times10^{10},$ the relative mass of neutral Hydrogen $\mu_{HI}=\frac{M_{HI}}{M_{sun}}\times10^{10}$ in
        which galactic total mass is $M_{HI}+M_{stellar}$}
          \begin{tabular}{c c c  c c c c c}
                $Galaxy $ & $type$  & $\mu_{star}$ & $\mu_{HI}$ & $\sigma$  & $r_c$ & $r_{qf}$  & $\varsigma$\\
                \hline
                NGC 247 & $LSB$  & $0.4$ & $0.13$ & $-1.71138\times10^7$ & $0.63315$ & $0.193396$ & $2$ \\
                \hline
                NGC 1003 & $LSB$  & $0.30$ & $0.82$ & $48381.3$  & $0.520752$ & $0.431666$  & $2$\\
                \hline
                NGC 2403 & $LSB$  & $1.1$ & $0.47$ & $1.70255$ & $0.470889$ & $0.416284$  & $2$\\
                \hline
                UGC 6930 & $LSB$   & $0.42$ & $0.31$ & $-4652.3$ & $0.393393$ & $0.269871$  & $2$ \\
                \hline

                UGC 6983 & $LSB$  & $0.57$ & $0.29$ & $7.75637\times10^{10}$ & $0.279081$ & $0.321622$   & $2$\\
                \hline
                NGC 300 & $HSB$  & $0.22$ & $0.13$ & $15468.$  & $0.979994$ & $0.155275$ & $1$ \\
                \hline
              NGC 3992 & $HSB$   &  $15.3$ &  $0.92$ & $9790.58$ & $-0.299233$ & $1.33564$ & $1$\\
                \hline
                NGC 4157 & $HSB$  &  $4.83$ & $0.79$ & $1.35773\times10^7$  & $0.149683$ & $0.793509$  & $1$\\
                \hline
               NGC 5907 & $HSB$  &  $9.7$ & $1.1$ & $3.89838\times10^7$  & $-0.873617$ & $1.26357$  & $1$\\
                \hline
                NGC 6503 & $HSB$  &  $0.83$ & $0.24$ & $20052.9$  & $0.0169275$ & $0.404241$   & $1$\\
                \hline
                NGC 6946 & $HSB$  & $2.7$ & $2.7$  & $3.2588$ & $0.196318$ & $0.931982$  & $1$ \\
                \hline
                  NGC 2903 & $HSB$  & $5.5$ & $0.31$ & $-11.4416$ & $0.198349$ & $0.726036$   & $1$\\
                \hline
            \end{tabular}
    \end{table}
    \begin{figure}[ht!] \centering
        \includegraphics[width=1.7in]{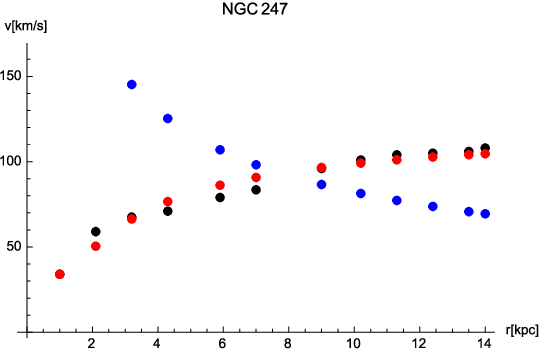}
        \includegraphics[width=1.7in]{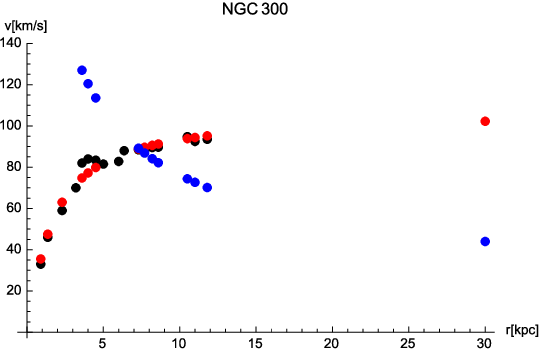}
        \includegraphics[width=1.7in]{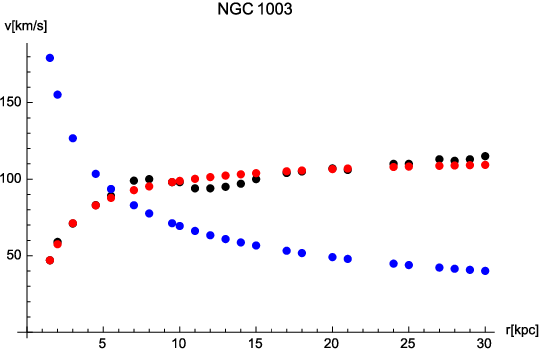}
        \includegraphics[width=1.7in]{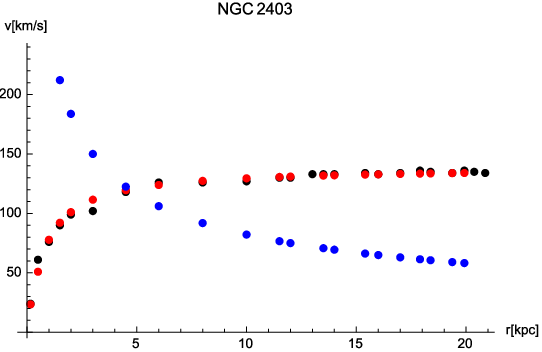}
        \includegraphics[width=1.7in]{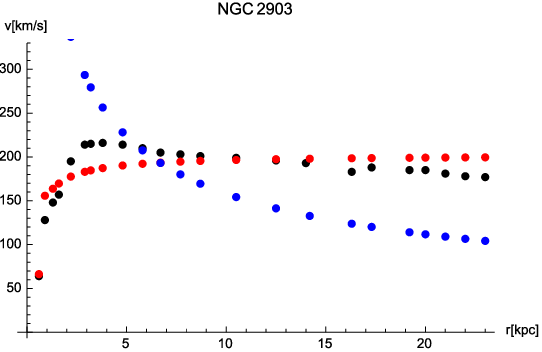}
        \includegraphics[width=1.7in]{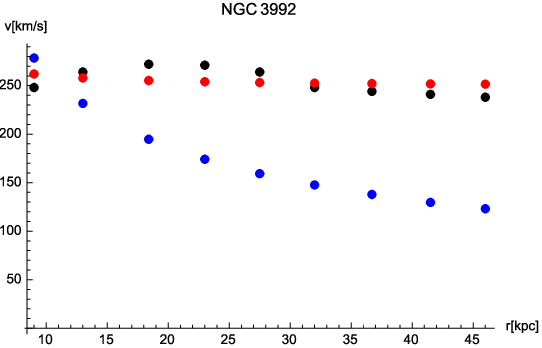}
        \includegraphics[width=1.7in]{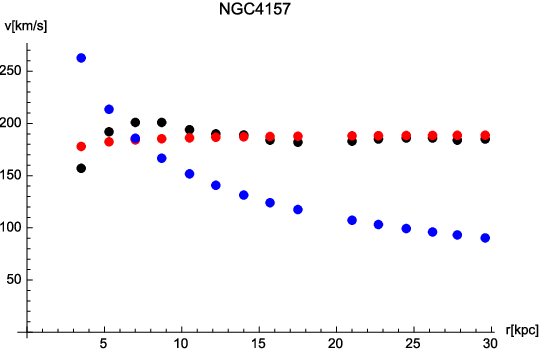}
        \includegraphics[width=1.7in]{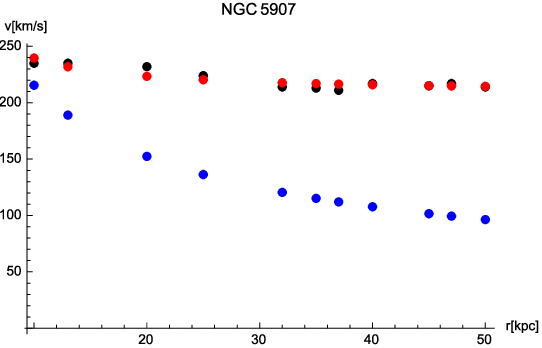}
        \includegraphics[width=1.7in]{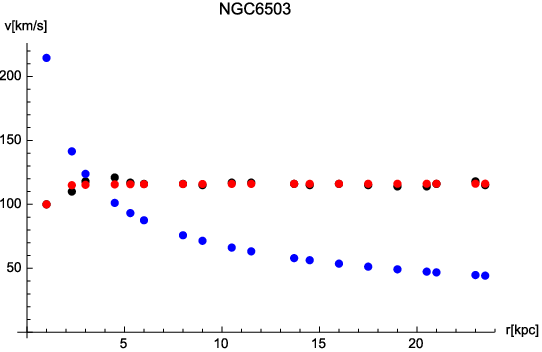}
        \includegraphics[width=1.7in]{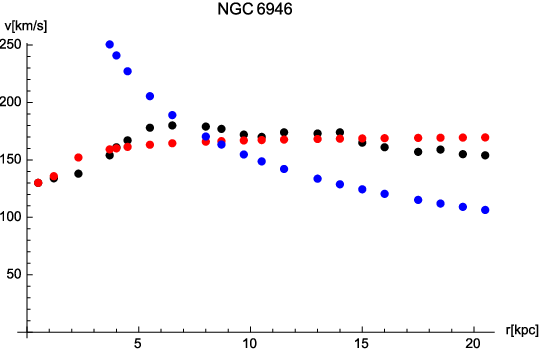}
        \includegraphics[width=1.7in]{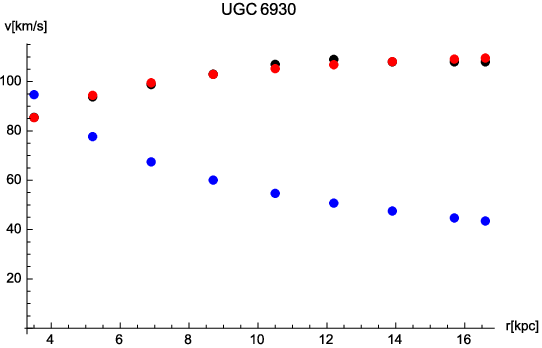}
         \includegraphics[width=1.7in]{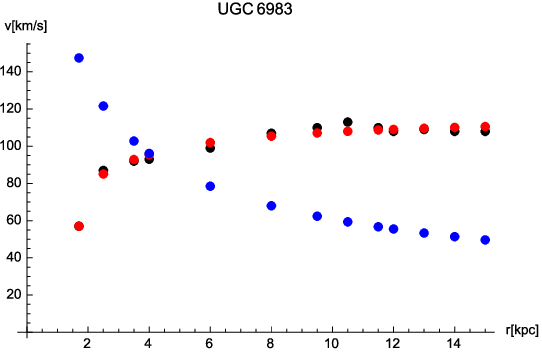}
        \caption{{\small Galaxy rotation curves for a set of 12
                galaxies with different size. Black-dots denote to observational
                data, red-dots denote to theoretical predictions of our model and blue-dots denote to the Newtonian counterpart of the galactic
                rotational curves. }}
    \end{figure}
   \subsection*{Baryonic Tully-Fisher relation}
The baryonic Tully-Fisher relation (BTF) is an empirical relation
between baryonic mass $M_B$
  and maximum rotation velocity $v_{max}$ which may be roughly expressed as
  \begin{equation}\label{MV}M \propto v_{max}^a\end{equation}
   where  $3.5\leq a\leq4$ is obtained from data analysis by applying the different gravity models such as MOG, MOND and etc. \cite{TFR}.
   Since we are interested in fitting rotation curves without any
   dark  matter halo, the baryonic mass of a galaxy contains
    stellar and gaseous components only such as $M_B=M_{star}+M_{gas}$. For galaxies whose rotation
    curves can be resolved, $v_{max}$ can be chosen as different forms for instance: the maximum observed velocity, the average
    velocity, or the velocity at a fixed radius where the rotation curves seams  flat. In ref. \cite{Fed} one can see different choices
    for $v_{max}$ which are applicable to study the empirical Tully-Fisher equation. Since the rotation curves are close to
     the flat region so
    all of these choices are effectively equivalent. In order to construct a baryonic Tully-Fisher relation, one
    needs to estimate the maximum rotation velocity
    of a galaxy. There are several ways in which this can be done.
    We choose $v_{max}$ to be asymptotically value of the equation
    (\ref{velocitydis}) at infinity  $v(r>>r_{qf}).$ To do so we substitute (\ref{rlarge}) into the relation (\ref{velocitydis})
    for which we obtain \begin{equation} v(r>>r_{qf})=v_{max}\bigg(\frac{r}{r_c+r}\bigg)^\frac{3\varsigma}{2}
    \end{equation} in which we defined \begin{equation}\label{vmax}v_{max}=\lim_{r\to\infty}v(r>>r_{qf})=\sqrt{\frac{4750}{\omega}}
    \sqrt{\frac{GM}{r_{qf}}}.\end{equation}
     In the following we intend to investigate whether the results of JBD-SVT theory is in general agreement with the BTF?.
     A log-log diagram of the maximum speed $v_{max}$ given by (\ref{vmax}) is plotted versus the baryonic mass
    of 12 observed galaxies in figure 2.
     The least-square fit to
  the data points is given by the following straight-line
  equation
\begin{equation}\label{TF}
\log\bigg(\frac{M}{M_{sun}}\times10^{10}\bigg)=a\log(v_{max})+b
\end{equation}
in which \begin{equation}\label{aLSB} a_{LSB}=4.13266,
~~~b_{LSB}=3.1236,\end{equation}
\begin{equation}\label{aHSB}a_{HSB}=4.19776,~~~b_{HSB}=2.77282\end{equation}  and
\begin{equation}\label{aLHSB} a_{LSB,HSB}=1.87782,~~~b_{LSB,HSB}=14.3757.\end{equation}
One can see that the equations (\ref{aLSB}) and (\ref{aHSB}) are
in a good agreement with BTF given by (\ref{MV}) alone and slope
of them are $4$ approximately, while a combined LSB $\&$ HSB given
by (\ref{aLHSB}) dose not satisfy the BTF and its slope reaches to
some small values with respect to $4.$ Difference of slopes with
respect to the prediction in ref. \cite{TFR} as $3.5<a<4$, maybe
resolved by regarding possibility observational errors, or by
using more samples of the observational data, regarding
cylindrical symmetry of the used spiral galaxies or by choosing
other samples for the maximum circular velocity defined in ref.
\cite{Fed}. Authors in most articles are used more than 100 sample
of observed galaxies data to obtain slope of the empirical Tully
Fisher relation.
 Since BTF is an empirical relation so the
existence of the Baryonic Tully-Fisher relation may implies that
the mass observed in baryons is the total mass \cite{mil} and it
challenges the dark matter hypothesis.
\begin{figure} \label{fig222}\centering
    \includegraphics[width=1.7in]{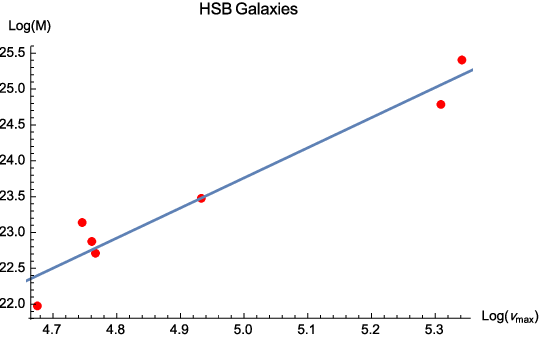}
     \includegraphics[width=1.7in]{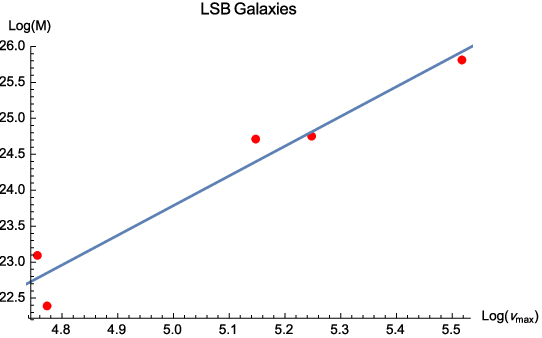}
      \includegraphics[width=1.7in]{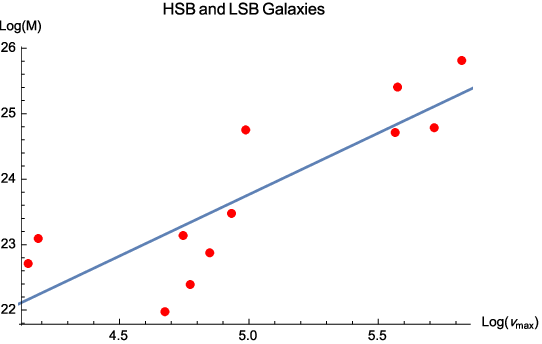}
    \caption{The baryonic Tully-Fisher relation of 12 spiral galaxy samples \cite{RH}. The rotation
    velocity at quasi flat region  $v_{qf}$
    is  given in the units of $km/s$ and mass is in units $10^{10} M_{sun}$. The red-dots are the results of
   our JBD-SVT model and they are obtained by calculating $v_{max}$ for observational data of 12 spiral galaxies. The
   blue line is obtained via a least-square fit to the observational data.}
\end{figure}
\section{Concluding remarks}
 The rotation curves of galaxies still remain as one of the profound challenges in the present day physics.
    In this paper we have considered an alternative view to the dark matter problem. In fact we used generalized Jordan-Brans-Dicke
scalar-vector-tensor (JBD-SVT) gravity  model  to generate galaxy
rotation curves where the time-like vector field $N_{\mu}$ is
coupled non-minimally to the BD scalar field
 $\phi$ and the background metric.
In weak field limits, we considered a power-law self-interaction
potential for the  vector field, which plays an important role to
produce a repulsive Yukawa like metric potential. In fact we
obtained that the vector field mass parameter is related to an
effective cosmological constant which supports the acceleration of
the universe in the large scalae structure. In other word we
obtained that in weak field limits metric around a galaxy under
consideration behaves as modified Schwarzschild de Sitter space
time. However this corrects galactic rotation curves by regarding
the experimental observational data. The results were surprisingly
the same as  ones which are given in \cite{BM} but provided their
physical basis are totally different.
 To test the
observational consequences of the circular velocities in our
model, we used the well-measured SPARC database \cite{SPARC} to
fit the theoretical rotation curves predicted by JBD-SVT to the
observational data. Our results are appropriately consistent with
the observations. We then demonstrate that our results are in a
good agreement with the baryonic Tully-Fisher relation
    for spiral galaxies. In short, the outlook of this work can be as follows:
    A
    vacuum sector of the Brans Dicke scalar tensor gravity in weak field limit which reduces to a Newtonian approach of the general relativity
     can not give a good fit to the galactic rotations curves (see blue dotes in figure 1).  While
    a time-like dynamical self interacting vector field moving on the curved background metric of a galaxy can produce a good
correspondence between theoretical results (see red-dotes in
figure 1) of galactic circular velocities and observational data
(see black dotes in figure 1). {On the other hand, baryonic Tully
Fisher relation confirms that one can describe galactic rotation
curves correctly without using unknown cold dark matter and just
with visible baryonic matter, using JBD-SVT theory.
 As further investigations one can study the solar system tests of the GBD-SVT theory, cosmological
implications of the theory such as domain walls and study the form of modified virial theorem in the theory.\\

\textbf{Acknowledgments}\\
The authors are grateful to the editor and anonymous referees for
their valuable comments and suggestions which cause to improve the
article for readers.


\begin{thebibliography}{99}
\bibitem{GB} G. Bertone, D. Hooper and J. Silk, Phys. Rep. \textbf{405}, 279 (2005).
\bibitem{Rub} V. C. Rubin, W. K. Jr. Ford, Astrophys. J. \textbf{159}, 379, (1970).
\bibitem{Ford} V. C. Rubin, E. M. Burbidge, G. R. Burbidge and K.
H. Prendergast, Astrophys. J. \text{141}, 885, (1965).
\bibitem{JM} J.W. Moffat, JCAP \textbf{0603}, 004 (2006); gr-qc/0506021.
\bibitem{RH} R. H. Sanders and S. S. Mc Gaugh, Ann. Rev. Astron.
Astrophys. \textbf{40}, 263 (2002); Astro-ph/0204521.
\bibitem{Bek} J. D. Bekenstein, Phys. Rev. D \textbf{71}, 069901 (2005).
\bibitem{MK} P. D. Mannheim, D. Kazanas, Astrophys. J. \textbf{342}, 635 (1989).
\bibitem{JCAP} J. W. Moffat, Phys. Lett. B  \textbf{355},447 (1995);   gr-qc/9411006
\bibitem{mash2007} B. Mashhoon, Annals Of Physics \textbf{16}, 57 (2007).
\bibitem{mash2009} B. Mashhoon, Phys. Lett. B \textbf{673}, 279 (2009).
\bibitem{MT} J. W. Moffat, V.T. Toth, Mon. Not. R. Astr. Soc. \textbf{395}, 25 (2009).
\bibitem{BM} J. R. Brownstein, J. W. Moffat, Astrophys. J. \textbf{636}, 721 (2006); Astro-ph/0506370.
\bibitem{BM2} J. R. Brownstein, J. W. Moffat, Mon. Not. R. Astr. Soc. \textbf{367} (2006).
\bibitem{MT2} J. W. Moffat, V.T. Toth, Astrophys. J. \textbf{680}, 1158 (2008).
\bibitem{Moffat1} J.W. Moffat, Found. Phys. \textbf{23}, 411 (1993).
\bibitem{Moffat2}J. W. Moffat Int. J. Mod. Phys. D \textbf{2}, 351 (1993).
\bibitem{CLJM}M. A Clayton, J. W. Moffat, Phys. Lett. B \textbf{460}, 263 (1999).
\bibitem{Barbero} J. F. G Barbero, E. J. S Villasenor, Phys. Rev. D \textbf{68}, 087501 (2003).
\bibitem{Jacob} T. Jacobson, D. Mattingly, Phys. Rev. D \textbf{64}, 024028 (2001).
\bibitem{Barbero2}J. F. Barbero, Phys. Rev. D \textbf{54}, 1492 (1996).
\bibitem{BD} C. Brans and R. Dicke, Phys. Rev. \textbf{124}, 925, (1961).
    \bibitem{GH1} H. Ghaffarnejad, Gen. Relativ. Gravit. \textbf{40}, 2229 (2008).
\bibitem{GH2} H. Ghaffarnejad, Gen. Relativ. Gravit. \textbf{41}, 2941 (2009).
\bibitem{GH3} H. Ghaffarnejad, Class. Quantum Grav. \textbf{27},
015008 (2010).
\bibitem{GH4} H. Ghaffarnejad, J. of Phys., \textbf{633}, 012020
(2015).
\bibitem{GH5} H. Ghaffarnejad and E. Yaraie, Gen. Relativ. Gravit.
\textbf{49}, 49 (2017).
\bibitem{GH6} H. Ghaffarnejad and H. Gholipour,  gr-qc/1706.02904
.
\bibitem{jac} T. Jacobson, D. Mattingly, Phys. Rev. D \textbf{64}, 024028, (2001).
\bibitem{CM1} C.M. Will, K. Nordvedt and Jr., Astrophys. J. \textbf{177}, 757 (1972).
\bibitem{CM2}K. Nordvedt, Jr. and C.M. Will, Astrophys. J. \textbf{177}, 775 (1972).
\bibitem{RW} R.W. Hellings, K. Nordvedt and Jr., Phys. Rev. D \textbf{7}, 3593 (1973).
\bibitem{MA} M.A. Clayton, J.W. Moffat, Phys. Lett. B \textbf{460}, 263 (1999).
\bibitem{MA2} M.A. Clayton, J.W. Moffat, Phys. Lett. B \textbf{477}, 269 (2000).
\bibitem{BA} B.A. Bassett, S. Liberati, C. Molina-Paris and M. Visser, Phys. Rev. D \textbf{62}, 103518 (2000).
\bibitem{CW1} C. M. Will, \textit{Theory and Experiment in Gravitational
    Physics,} Cambridge University press (1993), revised version:
arxiv:gr-qc/9811036.
\bibitem{CW2} C. M. Will, Living Rev. Rel. \textbf{9} (2006).http://www.livingreviews.org (2006).
\bibitem{EG} E. Gaztanaga and J. A. Lobo, Astrophys. J.
\textbf{548} 47 (2001).
\bibitem{RD} R. D. Reasenberg et al, Astrophys. J. \textbf{234},
925 (1961). \bibitem{DG1} D. Giannios, Phys. Rev. D71, 103511
(2005); gr-qc/0502122
\bibitem{Cho} Y. M. Cho, Phys. Rev. Lett. \textbf{68}, 3133
(1992).
\bibitem{weakBD} A. Barrosa, C. Romerob, Phys. Lett. A
\textbf{245}, 31 (1998); gr-qc/9712080 .
\bibitem{bookgal} J. Binney, S. Tremaine, \textit{Galactic dynamics}, 2nd edn. (Princeton
University Press, Princeton, 2008)
\bibitem{Blok} W. J. G. de Blok,
S. S. McGaugh and J. M. van der Hulst, Mon. Not. R. Astron. Soc.
\textbf{283}, 18 (1996).
\bibitem{SPARC} http://astroweb.case.edu/SPARC.
\bibitem{Sobuti} Y. Sobuti, Astronom. Astrophys., \textbf{464}, 921–925 (2007).
\bibitem{MRah} J. W. Moffat and S.
Rahvar, Mon. Not. R. Astron. Soc  \textbf{436}, 1439 (2013).
\bibitem{TFR} R. B. Tully and J. R. Fisher, Astron. Astrophys. \textbf{54}, 661 (1977).
\bibitem{Fed} F. Lelli, S. S.
McGaugh, J. M. Schombert and H. Desmond Mon. Not. R. Astron. Soc.
\textbf{201}, 1 (2019); astro-ph.GA/1901.05966


\bibitem{mil} M. Milgrom, Astrophys. J. \textbf{270}, 371 (1983).
\bibitem{SPARC1} F. Lelli, S. S.
McGaugh and J. M. Schombert, Astron. J. \textbf{152}, 157 (2016);
astro-ph.GA/1606.09251



\end{thebibliography}
\end{document}